\documentclass[]{spie}  

\usepackage{amsmath,amsfonts,amssymb}
\usepackage{graphicx}
\usepackage[colorlinks=true, allcolors=blue]{hyperref}

\title{The HERMES-TP/SP background and response simulations}

\author[a,b]{Riccardo~Campana}
\author[a,b]{Fabio~Fuschino}
\author[c,d]{Yuri~Evangelista}
\author[e]{Giuseppe~Dilillo}
\author[f]{Fabrizio~Fiore}

\affil[a]{INAF/OAS, Via Gobetti 101, I-40129, Bologna, Italy}
\affil[b]{INFN-Sezione di Bologna, Viale Berti Pichat 6/2, I-40127, Bologna, Italy}
\affil[c]{INAF/IAPS, Via Fosso del Cavaliere 100, I-00133, Roma, Italy}
\affil[d]{INFN-Sezione di Roma 2, Via della Ricerca Scientifica 1, I-00133, Roma, Italy}
\affil[e]{University of Udine, Via delle Scienze 206, I-33100, Udine Italy}
\affil[f]{INAF/OATS, Via G.B. Tiepolo 11, I-34131, Trieste, Italy}

\authorinfo{Further author information: Send correspondence to \url{riccardo.campana@inaf.it}}

\begin{document} 
\maketitle

\begin{abstract}
HERMES (\emph{High Energy Rapid Modular Ensemble of Satellites}) is an innovative mission aiming to observe transient high-energy events such as gamma-ray bursts (GRBs) through a constellation of CubeSats hosting a broadband X and gamma-ray detector. The detector is based on a solid-state Silicon Drift Detector (SDD) coupled to a scintillator crystal, and is sensitive in the 2 keV to 2 MeV band. An accurate evaluation of the foreseen in-orbit instrumental background is essential to assess the scientific performance of the experiment. An outline of the Monte Carlo simulations of the HERMES payload will be provided, describing the various contributions on the total background and the optimization strategies followed in the instrument design. Moreover, the simulations were used in order to derive the effective area and response matrices of the instrument, also as a function of the source location with respect to the detector frame of reference. 
\end{abstract}


\section{INTRODUCTION}
HERMES-Technologic and Scientific pathfinder (HERMES-TP/SP)\cite{fuschino19} is a constellation of six 3U nanosatellites hosting simple but innovative X-ray detectors for the monitoring of cosmic high energy transients, such as Gamma Ray Bursts and the electromagnetic counterparts of Gravitational Wave events. 
The main objective of HERMES-TP/SP is to prove that accurate position of high energy cosmic transients can be obtained using miniaturized hardware, with cost at least one order of magnitude smaller than that of conventional scientific space observatories and development time as short as a few years. 
The main goals of the projects are: 1) join the multimessenger revolution by providing a first mini-constellation for GRB localizations with a total of six units, performing a first experiment of GRB triangulation with miniaturized instrumentation; 2) develop miniaturized payload technology for breakthrough science; 3) demonstrate commercial, off-the shelf (COTS) components applicability to challenging missions, contribute to the so-called Space 4.0 goals; push and prepare for high reliability large constellations.

The HERMES-TP project is funded by the Italian Ministry for education, university and research and the Italian Space Agency. The HERMES-SP project is funded by the European Union’s Horizon 2020 Research and Innovation Programme under Grant Agreement No. 821896. 
The constellation should be tested in orbit in 2022. 
HERMES-TP/SP is intrinsically a \emph{modular} experiment that can be naturally expanded to provide a global, sensitive all sky monitor for high energy transients.

The foreseen detector for this experiment employs the solid-state Silicon Drift Detectors (SDD) developed by INFN and FBK in the framework of the ReDSoX Collaboration\footnote{\url{http://redsox.iasfbo.inaf.it}}. These devices, being sensitive to both X-ray and optical photons, and characterised by a very low intrinsic electronic noise, and can be exploited in the so-called ``siswich'' architecture, acting both as direct X-ray detectors (the operative X-mode) and as photodetectors for the scintillation light produced by the absorption of a gamma-ray in an inorganic scintillator crystal (S-mode). This allows for the realisation of a single, compact experiment with a sensitivity band from a few keV up to a few MeVs for X and gamma-rays, and with a high temporal resolution ($<$$\mu$s). 

\begin{figure}[htbp]
\centering
\includegraphics[width=12cm]{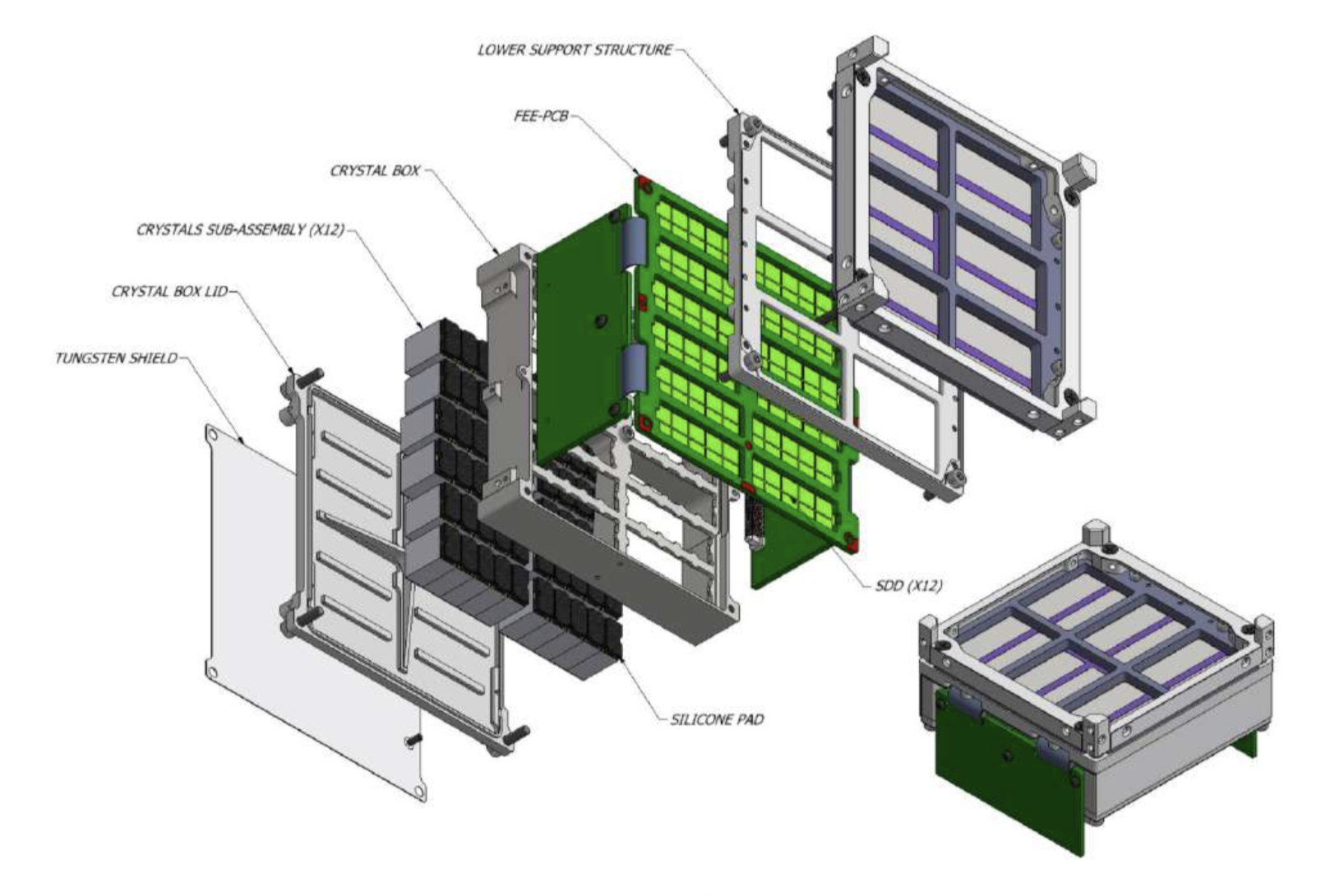}
\caption{Exploded view of the HERMES detector.}
\label{f:pl_sketch}
\end{figure}

Figure~\ref{f:pl_sketch} shows a sketch of the proposed payload (which will fit in a standard 3U CubeSat platform). 
Each of the sixty $12.10 \times 6.94 \times 15.00$ mm$^3$ scintillation crystals (made of cerium-doped gadolinium-aluminium-gallium garnet, GAGG:Ce) is coupled to two SDD channels. Ten SDD channels are integrated in a single monolithic 2$\times$5 matrix. The crystals are contained in a stainless steel box, shielded with a tungsten layer.
A more extensive description of the payload and of the present development status is given in other proceedings in this volume\cite{fiore20,evangelista20}.

Since the nanosatellites will be launched as secondary payloads in a low-Earth orbit, they are subject to a relatively large amount of high-energy radiation fluxes (mostly primary, secondary and trapped charged particles, and diffuse photon backgrounds), which are the main sources of the on-orbit scientific background. In this paper the results of Monte Carlo simulations of the present design of the HERMES payload in this environment are shown, discussing the performance and response of the instrument. 

\section{BACKGROUND SOURCES}\label{s:sources}
\begin{figure}[htbp]
\centering
\includegraphics[width=12cm]{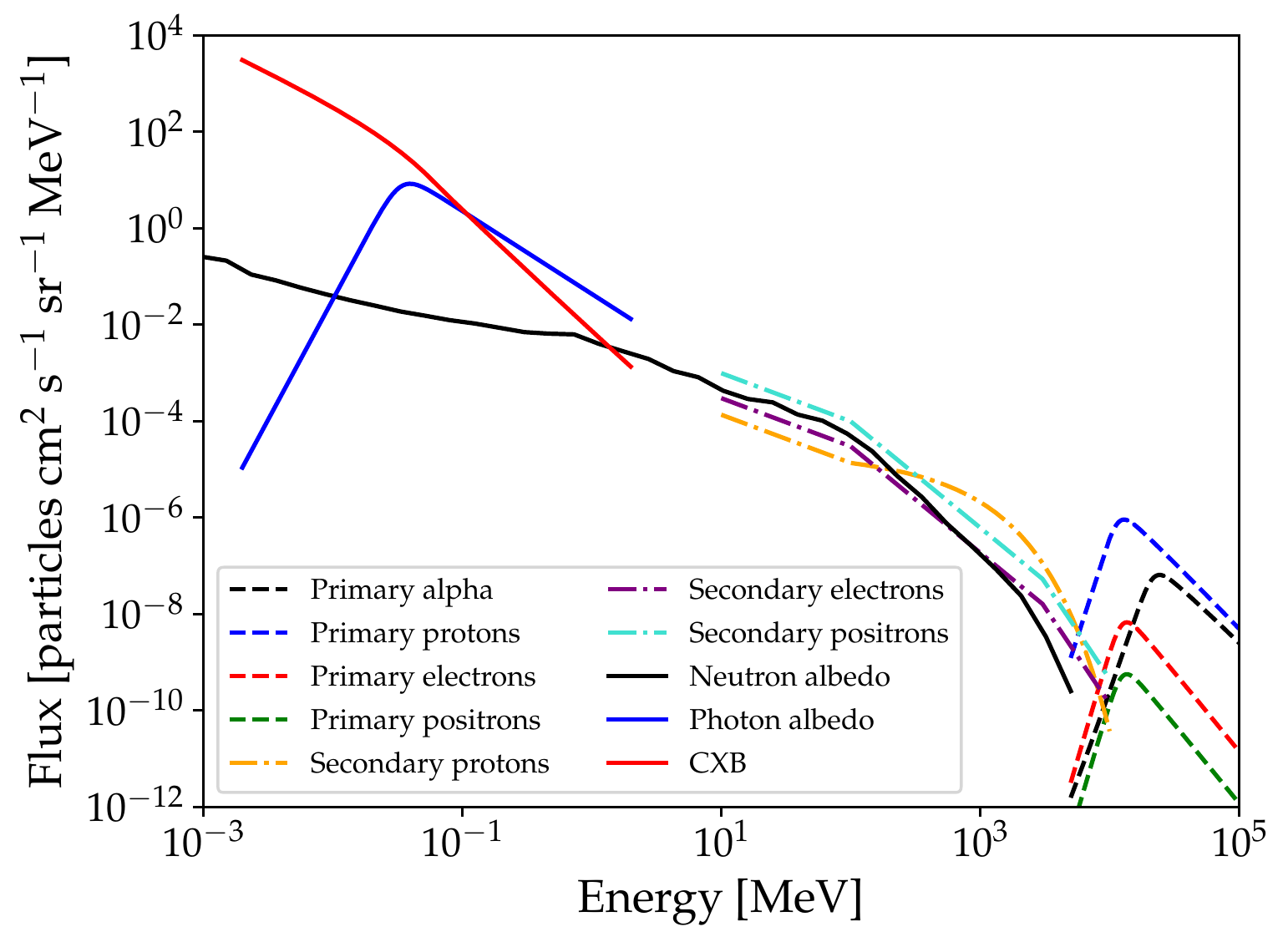}
\caption{Contributions to the HERMES scientific background.}
\label{f:bkg_src}
\end{figure}

The background environment in an equatorial, low-Earth orbit (LEO), has been extensively discussed in detail elsewhere\cite{campana13}. In this section the main populations of particles and photons which contribute to the on-board scientific background, and which are the inputs to the Monte Carlo simulations of the HERMES payload, are summarized.  Figure~\ref{f:bkg_src} shows the differential spectrum of each of these contributions.

\subsection{Primary cosmic rays}

The unmodulated value of the cosmic ray proton spectrum is given by the BESS and AMS measurements\cite{alcaraz00a, mizuno04}. We assume the primary proton spectrum for solar minimum conditions.  
The fraction of cosmic ray positrons to electrons is found to be rather independent of the energy, i.e. the spectrum of primary positrons has the same slope of the electron one, but a different normalization.
Furthermore, we consider only the cosmic ray primary Helium nuclei (alpha particles),  because the contribution of heavier nuclei is in the same range of the uncertainties in the primary flux. 

For all these components, we follow the analytical model of Mizuno et al.\cite{mizuno04} for low geomagnetic latitudes (Figure~\ref{f:bkg_src}).

\subsection{Secondary particles}
For the low altitude equatorial Earth orbits considered, the impinging proton spectrum (outside the trapped particle belt, i.e. the South Atlantic Anomaly) consists, beside the primary component discussed before, also of a secondary, quasi-trapped component, originating from and impacting to the Earth atmosphere (sometimes in the literature they are referred as the “splash” and “reentrant” components). 
AMS measurements\cite{alcaraz00a} showed that this secondary component is composed by a short-lived and a long-lived particle population, both originating from the regions near the geomagnetic equator.
For low geomagnetic altitudes, the modelling the secondary equatorial proton and lepton spectrum is as a cutoff power-law or a broken power law\cite{campana13, mizuno04} (Figure~\ref{f:bkg_src}).
At variance with respect to the primary particles, in the geomagnetic equatorial region the positrons are predominant with respect to the electrons. The spectrum has the same shape, but the ratio e+/e- is about 3.3.

As reported by the official ESA ECSS documents\footnote{\url{http://space-env.esa.int/index.php/ECSS-10-4.html}}, there is presently no model for atmospheric albedo neutron fluxes considered mature enough to be used as a standard. To account for the flux of neutrons produced by cosmic-ray interactions in the Earth atmosphere, we used the results of the Monte Carlo radiation transport code based QinetiQ Atmospheric Radiation Model (QARM\cite{lei04,campana13}, Figure~\ref{f:bkg_src}).

\subsection{Photon background}

For the cosmic X-ray and $\gamma$-ray diffuse background we assume the Gruber et al.\cite{gruber99} analytic form, derived from HEAO-1 A4 measurements, valid in the range from 3 keV to 100 GeV (Figure~\ref{f:bkg_src}).

The secondary photon background is due to the cosmic-ray (proton and leptonic components) interaction with the Earth atmosphere. As such, it has a strong zenith dependence and a higher flux, for unit of solid angle, than the CXB for energies above 70 keV. 
We assume the parameterized function for the albedo spectrum as given by Swift/BAT\cite{ajello08} measurements (Figure~\ref{f:bkg_src}), that agrees in the range above 50 keV, after some corrections, with previous measurements\cite{campana13}.

\section{THE MONTECARLO MASS MODEL}
\begin{figure}[htbp]
\centering
\includegraphics[width=10cm]{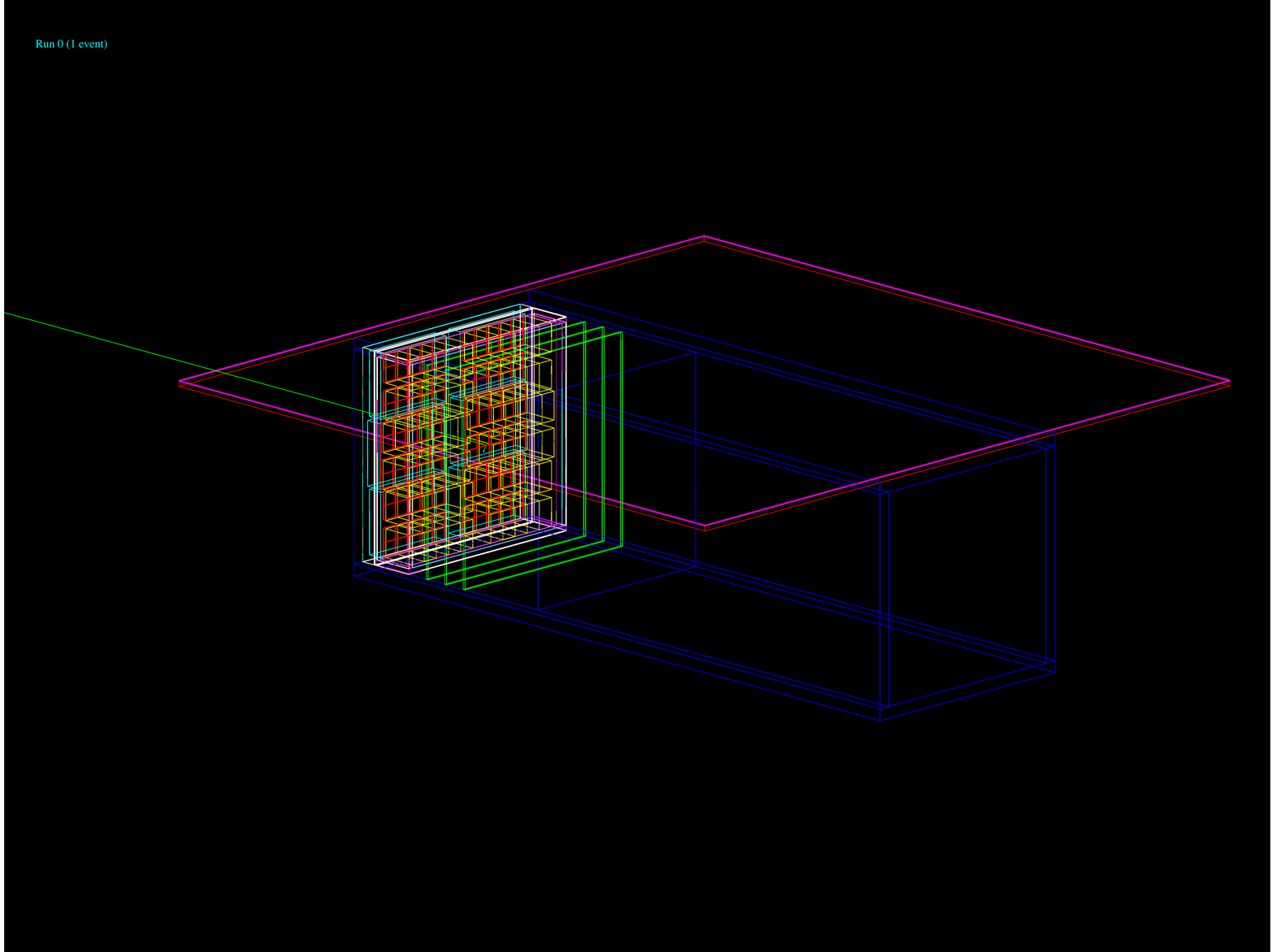}
\caption{The HERMES Geant4 Monte Carlo mass model.}
\label{f:massmodel}
\end{figure}

The full HERMES payload mass model, implemented in Geant4 \cite{agostinelli03}, is shown in Figure~\ref{f:massmodel}. The mass model includes all the significant structures surrounding the SDD detectors. From top to bottom, with respect to the detector boresight, we have the following elements:
\begin{enumerate}
\item	Multi-layer insulation (Kapton + Aluminum)
\item	Detector support structure (Stainless steel)
\item Optical filter (Kapton + Aluminum)
\item	SDDs (Silicon)
\item	Optical coupler (Silicone)
\item	Crystals (GAGG:Ce)
\item	Crystal box (Stainless steel with tungsten shielding on two sides and bottom)
\item	Printed circuit boards for back-end electronics, power supply units and payload data handling unit (FR4)
\item	Simplified spacecraft bus and solar panel structure (Aluminum with effective density)
\end{enumerate}

In other words, the payload structures are described with reasonable accuracy (in particular for the assumed materials and thicknesses) while the satellite bus structures are representative of the total mass but lack precise geometrical details. The mass model will be updated in the future with a more accurate description of the latter components.

The input sources (Section~\ref{s:sources}) are simulated isotropically on a sphere surrounding the CubeSat mass model, and the energy deposits on both the SDDs and the scintillator crystals are recorded. The event reconstruction then takes into account the noise of the detector front-end electronics and the optical coupling between detector and crystals. As a result, separate event lists for both X-mode (direct photon detection on the SDD) and S-mode (photon interaction in the crystal and subsequent scintillation light detection) are generated.

\section{RESULTS}

\subsection{Background}
The following Figure \ref{f:totalbkg} reports the total simulated spectra for both X-mode and S-mode. The CXB is the dominant contribution to the background in X-mode, due also to the large field of view. Photon backgrounds (CXB and Earth albedo) likewise dominate the S-mode background up to energies above 500 keV, where the particle-induced background becomes dominant. The S-mode background shows also an evident 511 keV annihilation line that could be used for calibration purposes.

\begin{figure}[htbp]
\centering
\includegraphics[width=10cm]{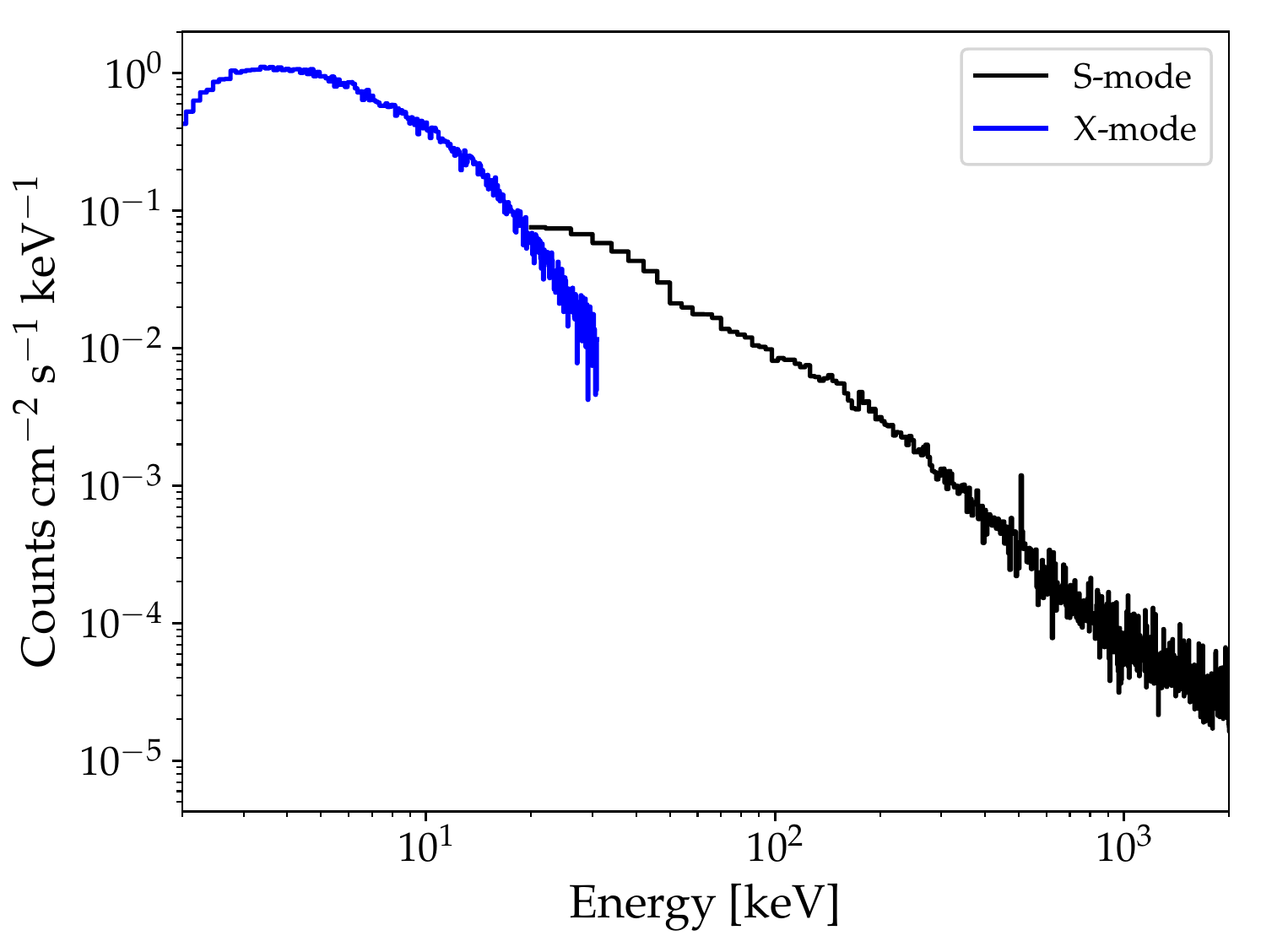}
\caption{The HERMES simulated in-orbit background.}
\label{f:totalbkg}
\end{figure}

The predicted count rate in various energy bands is shown in the following Table~\ref{t:bkgrates}.

\begin{table}[htp]
\caption{Predicted background count rates in various energy bands. The particle contribution includes primary and secondary protons, alpha particles and leptons,  and albedo neutrons.}
\begin{center}
\begin{tabular}{lccccc}
\hline
Contribution &	3--20 keV & 20--50 keV & 50--300 keV & 300--2000 keV & 3--2000 keV \\
& [counts s$^{-1}$] & [counts s$^{-1}$] & [counts s$^{-1}$] & [counts s$^{-1}$] & [counts s$^{-1}$] \\
& X-mode & both modes & S-mode & S-mode & both modes \\
\hline
CXB	& 501.7 & 94.0 & 46.8 & 3.6 & 646 \\
Earth $\gamma$ albedo & 0.3 & 8.6 & 22.9 & 6.2 & 38\\
Particles	& 0.03 & 0.39 & 2.3 & 4.2 & 7\\
\hline
Total & 503 & 103 & 72 & 14 & 692 \\
\hline
\end{tabular}
\end{center}
\label{t:bkgrates}
\end{table}%

\subsection{Strategies for background reduction}
The fine segmentation of the HERMES detector (60 individual crystals readout by 120 SDD channels) allows also to efficiently recognize and reject particle-induced background events, without affecting the photon-induced events.
Usually, particles incident on the HERMES detector will leave ionization streaks intersecting many crystals simultaneously. Moreover, often the energy deposit on a GAGG:Ce crystal is large (e.g. for particles at the minimum of ionization), i.e. above the 2 MeV sensitivity band limit. Therefore, an effective suppression for particle-induced events can be obtained by an hardware or software-implemented onboard filter which recognizes and rejects all events having either:
\begin{itemize}
\item	An energy deposit in a single crystal above 2 MeV (saturated events)
\item	A simultaneous trigger of at least 3 crystals (i.e. 6 SDD channels).
\end{itemize}

The efficiency of such a filter has been evaluated using the Monte Carlo mass model and the input sources discussed above. The following Table~\ref{t:rejection} reports the overall event rejection efficiency. It is evident how such a filter is able to reject most (up to 94\%, depending on particle type and spectrum) of the incident particle population, while leaving intact the events due to the interaction of photons.

\begin{table}[htp]
\caption{Rejection efficiency $\varepsilon_\mathrm{rej}$ using the event multiplicity criteria discussed in the main text, for the various background sources, as calculated through Monte Carlo simulations.}
\begin{center}
\begin{tabular}{ll}
\hline
Contribution &	$\varepsilon_\mathrm{rej}$ \\
\hline
Primary $\alpha$		&	$0.91$ \\
Primary $p^+$			&	$0.92 $\\
Primary $e^+$			& 	$0.90 $\\
Primary $e^-$			&	$0.90 $\\
Secondary $p^+$		&	$0.94 $\\
Secondary $e^+$		&	$0.78 $\\
Secondary $e^-$		&	$0.80 $\\
Neutron albedo			&	$0.15 $\\
Photon albedo			&	$1\times10^{-3}$ \\
CXB					&	$5\times10^{-5}$ \\
\hline
\end{tabular}
\end{center}
\label{t:rejection}
\end{table}%

Moreover, the Monte Carlo simulator, thanks to the possibility to specify at runtime the various materials and thicknesses, has been used to iteratively refine the final design of the detector, by studying the various trade-offs between the possible materials for the crystal box, taking into account their mass, mechanical and thermal behaviour and their shielding properties. In particular, the tungsten layer thickness on the bottom and lateral sides of the box has been chosen to effectively filter the photon component of the background (in particular the Earth albedo $\gamma$-rays) up to an energy of a few hundreds of keV.

\begin{figure}[htbp]
\centering
\includegraphics[width=7cm]{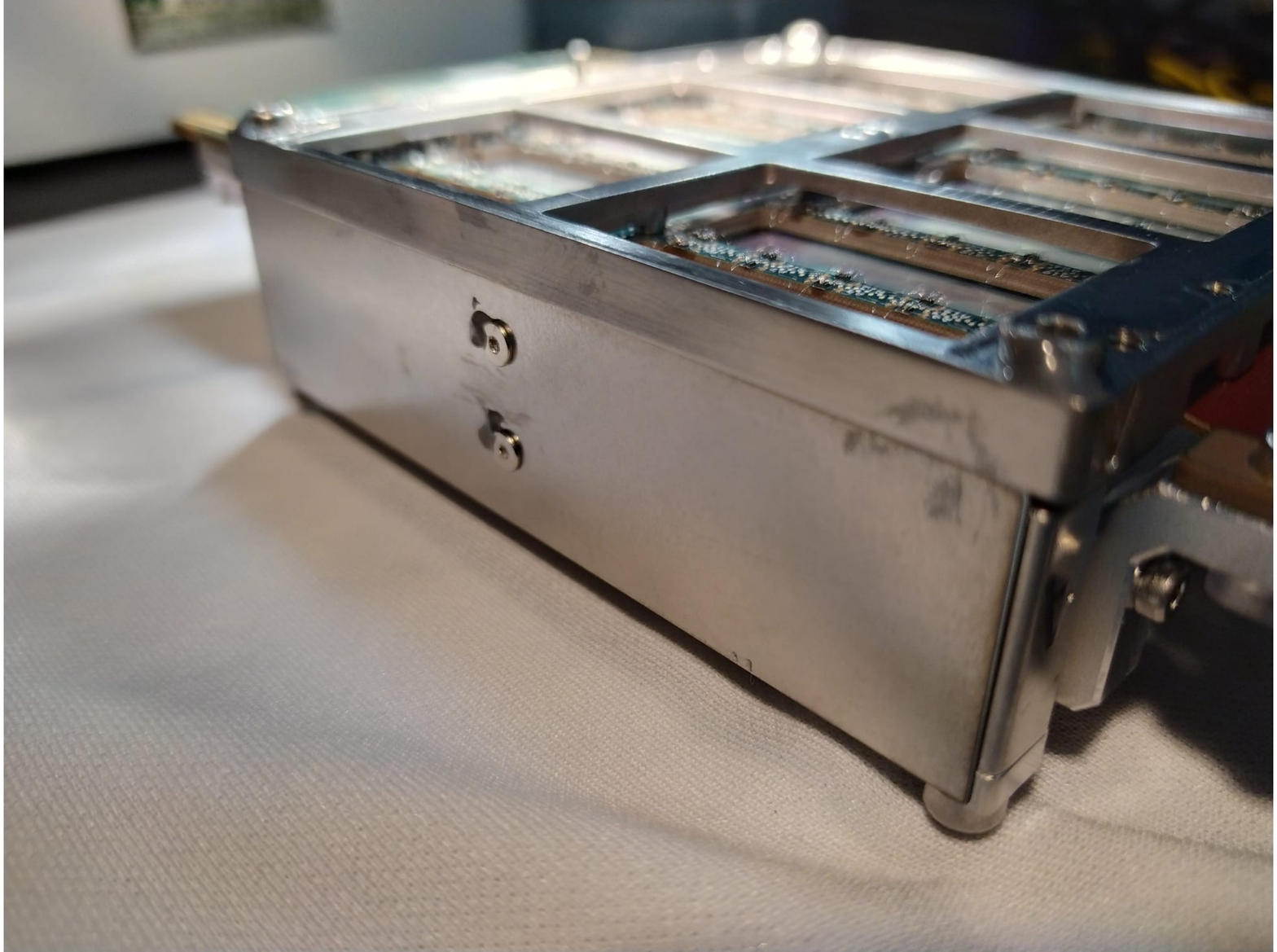} \includegraphics[width=7cm]{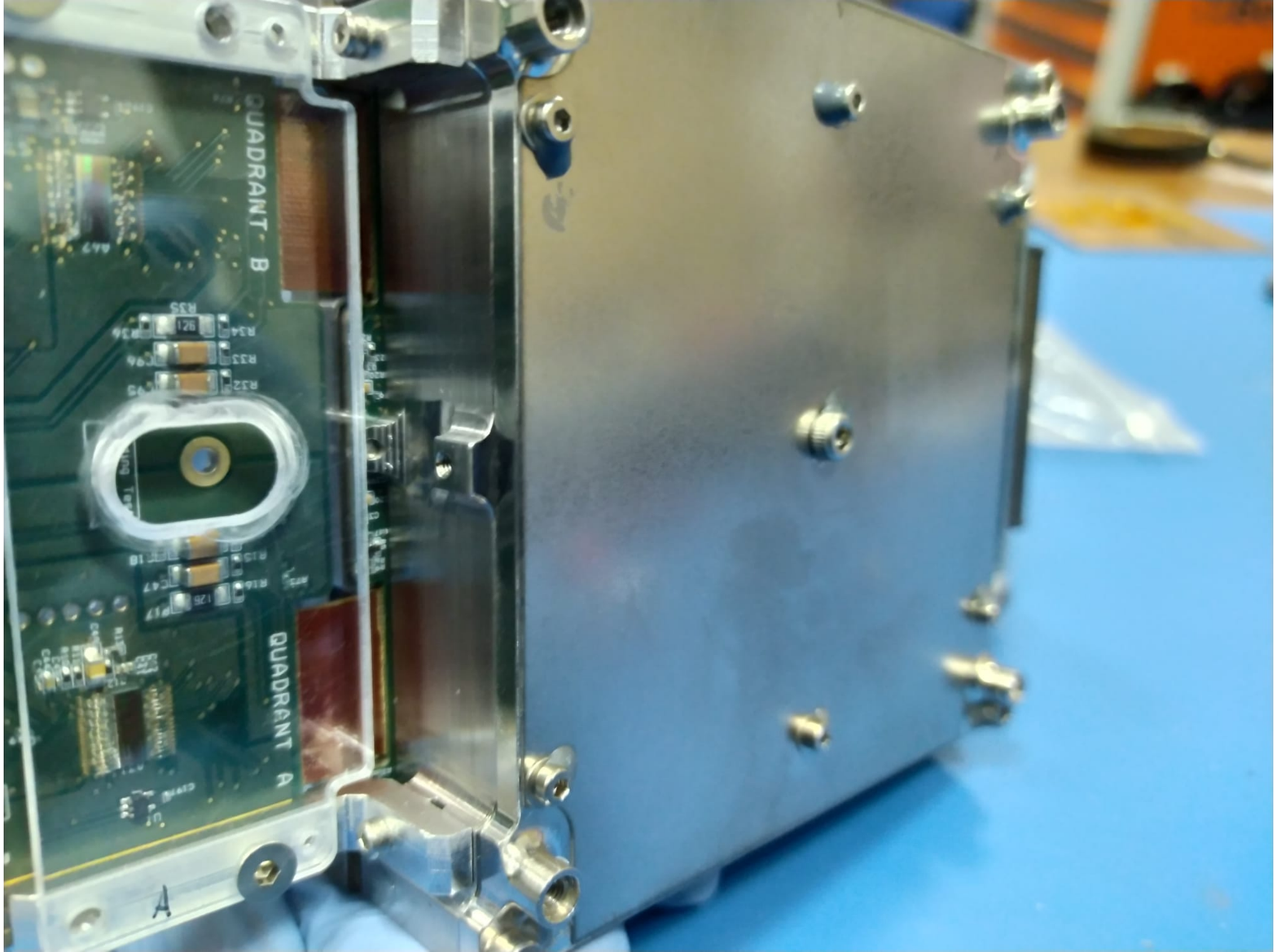}
\caption{Photos of the HERMES Demonstration Model (DM) showing the side (left panel) and bottom (right panel) tungsten shielding layer.}
\label{f:shielding}
\end{figure}

Figure~\ref{f:shielding} shows the HERMES Demonstration Model, integrated in Summer 2020\cite{evangelista20}, in which is visible the final W 200~$\mu$m thick shielding, on two sides of the crystal box and on the bottom side.

\subsection{Response}
\begin{figure}[htbp]
\centering
\includegraphics[width=10cm]{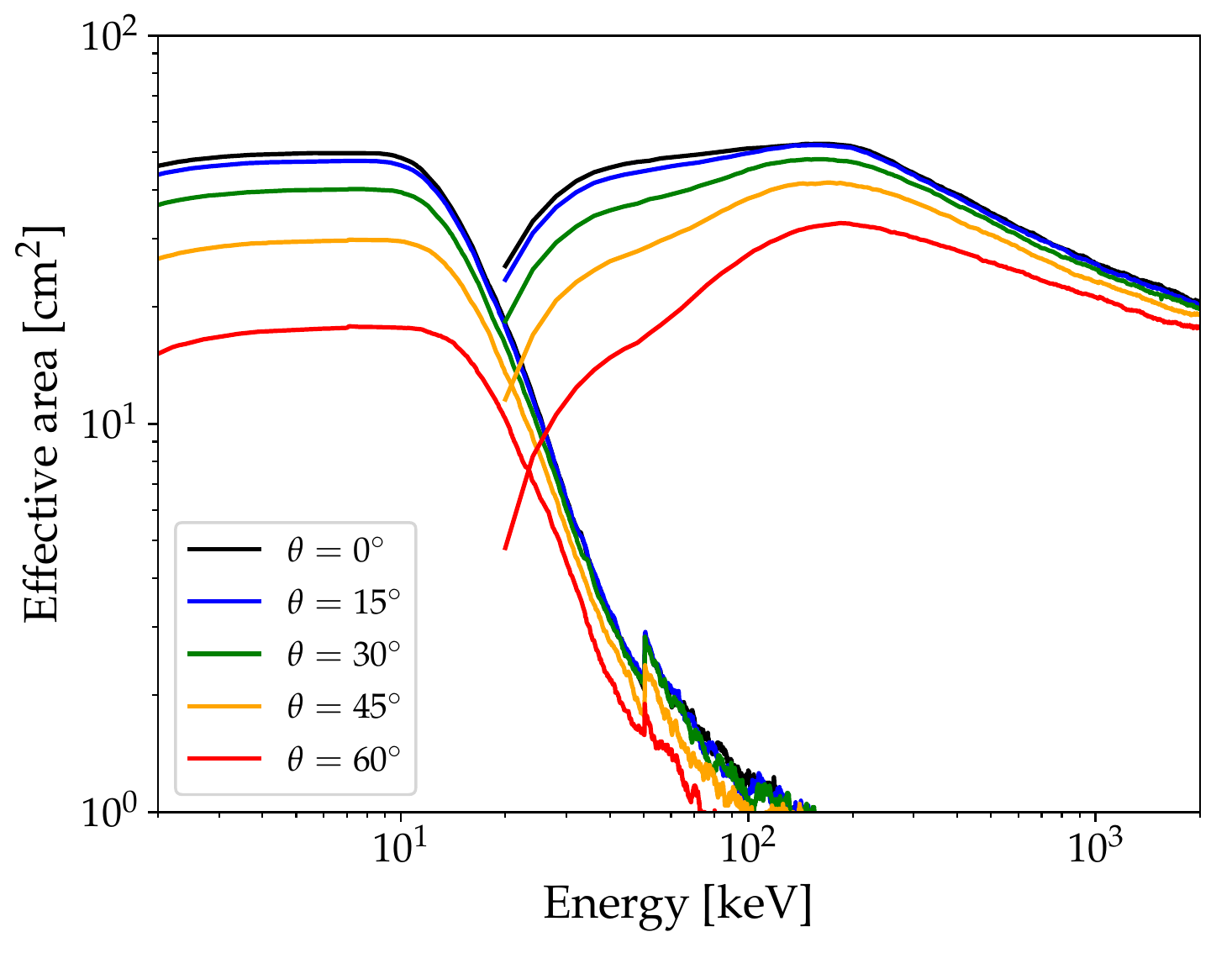}
\caption{The HERMES effective area for various off-axis angles. The leftmost curves show the X-mode effective area, while the rightmost the S-mode area.}
\label{f:effarea_offaxis}
\end{figure}

The HERMES mass model can be also used to derive the instrumental response functions, useful for the scientific performance simulations. Using monochromatic parallel beams of photons incident on-axis and at different off-axis angles the \emph{redistribution function} (RMF) and the effective area as a function of the energy can be generated for both X-mode and S-mode. Figure~\ref{f:effarea_offaxis} shows the effective areas as a function of the off-axis angle.

\subsection{Expected orbital background modulation}
HERMES foreseen orbit is at low inclination, thus spanning a restricted range of geomagnetic latitudes and grazing the South Atlantic Anomaly only in its outermost regions. 
Therefore, the expected modulation along the orbit of the background count rate is expected to be small and to depend mostly on the detector attitude (i.e. on the angle between the pointing direction and the Earth location in the sky) and on the residual variation of geomagnetic field characteristics (e.g. vertical cutoff rigidity\cite{campana14}). Preliminary simulations show a modulation up to a few tens of percent, depending on the pointing strategy (e.g. zenith vs. fixed pointing, boresight declination, etc.). 

\begin{figure}[htbp]
\centering
\includegraphics[width=5cm,height=5cm]{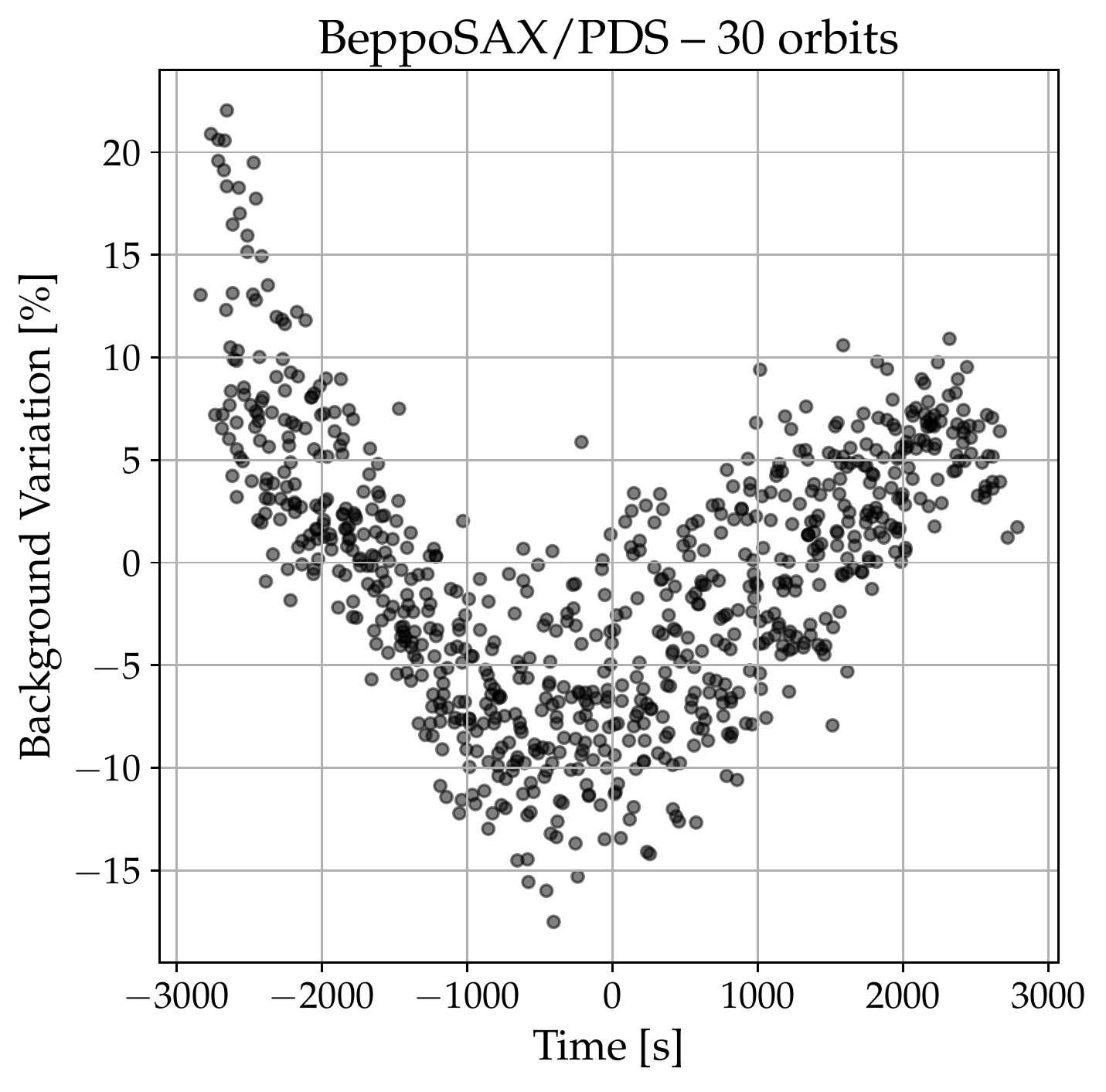}%
\includegraphics[width=5cm,height=5cm]{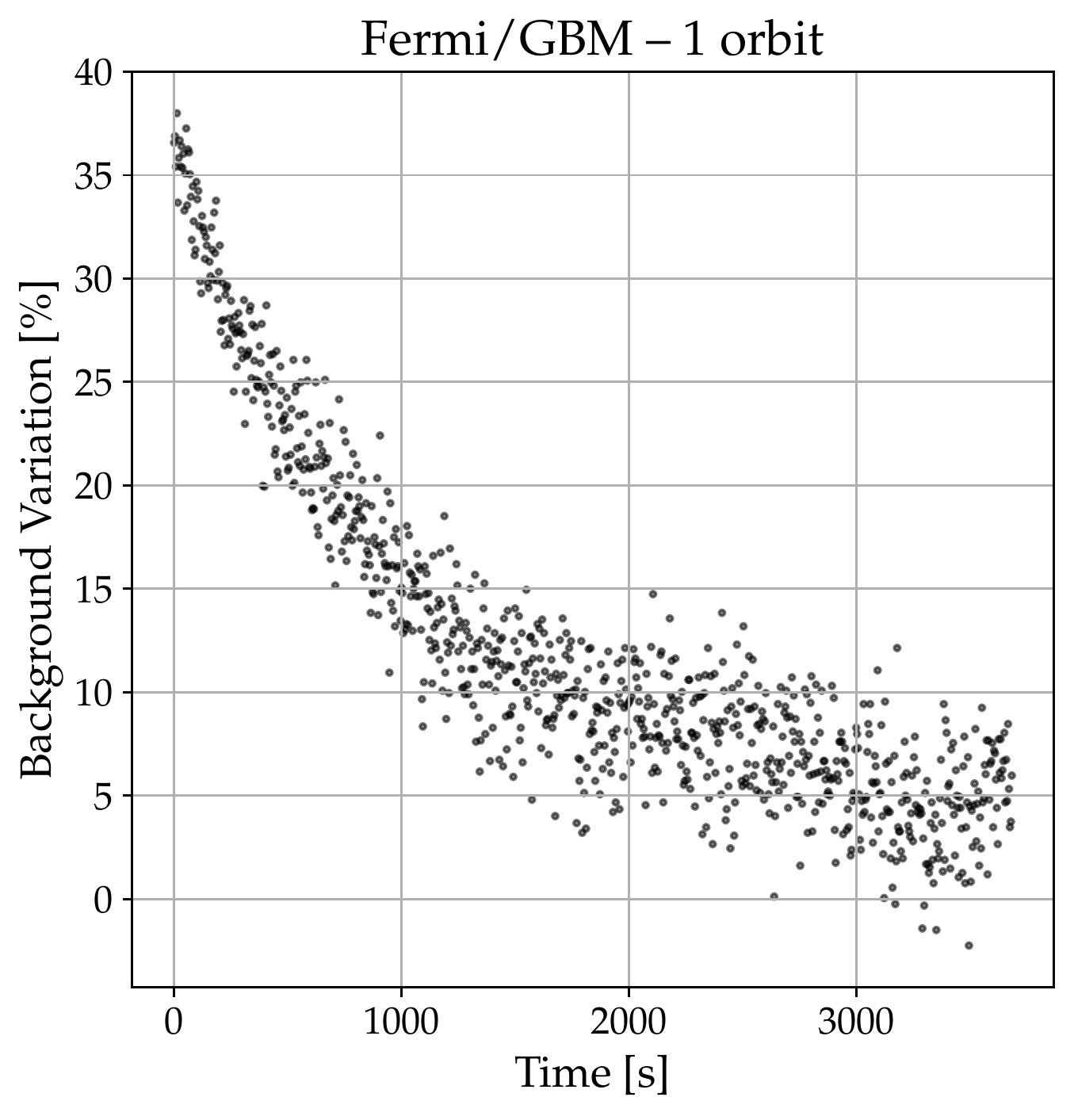}%
\includegraphics[width=5cm,height=5cm]{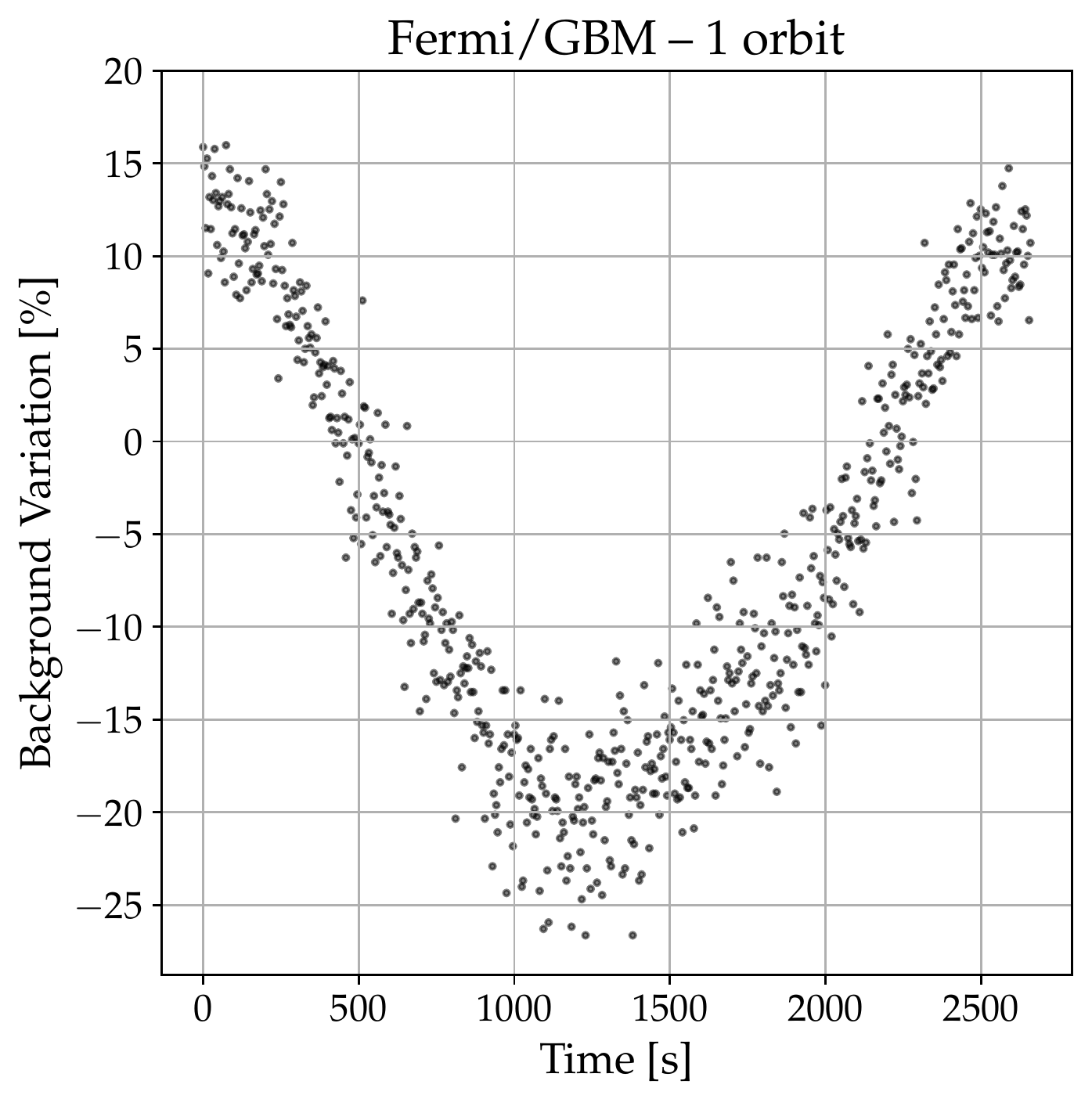}
\caption{
\emph{Left panel}: Background variation for an example BeppoSAX/PDS observation (flying in a $\sim$600 km altitude, 3.9$^\circ$ inclination orbit), for 30 successive orbits, folded with respect to the orbital phase. The data points are normalized with respect to the average count rate along this time span. \emph{Central panel}: Background rate variation for an example orbit for Fermi/GBM (550~km altitude, 25.6$^\circ$ inclination), just after a deep SAA transit. \emph{Right panel}: a different Fermi/GBM orbit, not transiting the SAA.
}
\label{f:bkgmod}
\end{figure}

It is instructive to compare the background rate orbital modulation measured by similar high energy instruments, such as BeppoSAX/PDS and Fermi/GBM.
Figure~\ref{f:bkgmod} show the background rate as a function of the orbital phase. It is apparent how for these instruments the background fluctuations are limited to a maximum of $\sim$20\% of the average background count rate, on a $\sim$90~min timescale. A similar behavior is expected also for HERMES.

\section{Conclusions}

In the framework of the HERMES-TP/SP project, a mass model and simulator for its payload has been implemented using the Geant-4 toolkit. Together with a model of the radiative environment in a LEO, it allowed to obtain instrumental background and instrument response files. 

Future developments of the mass model, given the maturity of the instrument design and of the satellite platform development, would include a more detailed mass model for the spacecraft bus and higher level of details around the detectors. Moreover, the background simulations should include a study of activation-induced background (given the equatorial LEO, preliminary evaluations shows that this should be a minor, albeit non-negligible, contribution to the overall background). The detailed study of event topology should lead to more efficient background rejection filters, thus optimizing the overall sensitivity of the instrument.

\acknowledgments 
This project has received funding from the European Union Horizon 2020 Research and Innovation Framework Programme under grant agreement HERMES-Scientific Pathfinder n. 821896 and from ASI-INAF Accordo Attuativo HERMES Technologic Pathfinder n. 2018-10-HH.0.

\bibliography{report} 
\bibliographystyle{spiebib} 

\end{document}